\documentclass[prb,amsmath,amssymb,superscriptaddress,twocolumn]{revtex4}
\usepackage{float}
\usepackage{amsmath}
\usepackage{amssymb}
\usepackage{amsfonts}
\usepackage{euscript}
\usepackage{enumerate}
\usepackage{hhline}
\usepackage{pslatex}
\usepackage{tabularx}
\usepackage[usenames,dvipsnames]{xcolor}

\usepackage{graphicx}

\usepackage{dcolumn}
\usepackage{bm}
\usepackage[sort&compress]{natbib}

\usepackage{lipsum}

\makeatletter
\renewcommand{\@biblabel}[1]{#1. }
\renewcommand{\@dotsep}{500}
\renewcommand{\@pnumwidth}{0em}
\renewcommand{\l@figure}[2]{
\@dottedtocline{1}{1.5em}{2em}{Figure #1}{}\vspace{15pt}}

\newcommand{\ks}[1]{\textcolor{black}{#1}}
\newcommand{\kst}[1]{\textcolor{black}{#1}}
\newcommand{\xl}[1]{\textcolor{black}{#1}}
\newcommand{\xlt}[1]{\textcolor{black}{#1}}

\usepackage[normalem]{ulem}

\begin{document}
\title{High-$Q$ slow light \xlt{and its localization in a} photonic crystal microring}

\author{Xiyuan Lu}\email{xiyuan.lu@nist.gov}
\affiliation{Microsystems and Nanotechnology Division, Physical Measurement Laboratory, National Institute of Standards and Technology, Gaithersburg, MD 20899, USA}
\affiliation{Institute for Research in Electronics and Applied Physics and Maryland NanoCenter, University of Maryland,
College Park, MD 20742, USA}
\author{Andrew McClung}
\affiliation{Department of Electrical and Computer Engineering, University of
Massachusetts Amherst, Amherst, MA 01003, USA}
\author{Kartik Srinivasan} \email{kartik.srinivasan@nist.gov}
\affiliation{Microsystems and Nanotechnology Division, Physical Measurement Laboratory, National Institute of Standards and Technology, Gaithersburg, MD 20899, USA}
\affiliation{Joint Quantum Institute, NIST/University of Maryland, College Park, MD 20742, USA}

\date{\today}

\begin{abstract}
\noindent We introduce a photonic crystal ring cavity \kst{that resembles an internal gear and unites} photonic crystal (PhC) and whispering gallery mode (WGM) concepts. \xlt{This `microgear' photonic crystal ring (MPhCR)} is created by applying a periodic modulation to the inside boundary of a microring resonator to open a large bandgap, as in a PhC cavity, while maintaining the ring's circularly symmetric outside boundary and high quality factor ($Q$), as in a WGM cavity. \kst{The MPhCR} targets a specific WGM to open a large PhC bandgap up to tens of free spectral ranges, compressing the mode spectrum while maintaining the high-$Q$, angular momenta, and waveguide coupling properties of the WGM modes. In particular, near the dielectric band-edge, we observe modes whose group velocity is slowed down by 10 times relative to conventional microring modes while supporting $Q~=~(1.1\pm0.1)\times10^6$. This $Q$ is $\approx$~50$\times$ that of the previous record in slow light devices. Using the slow light design as a starting point, we further demonstrate the ability to localize WGMs into photonic crystal defect (dPhC) modes for the first time, enabling a more than 10$\times$ reduction of mode volume compared to conventional WGMs while maintaining high-$Q$ up to (5.6$\pm$0.1)$\times$10$^5$. Importantly, this additional dPhC localization is achievable without requiring detailed electromagnetic design. Moreover, controlling their frequencies and waveguide coupling is straightforward in \kst{the MPhCR}, thanks to its WGM heritage. By using a PhC to strongly modify fundamental properties of WGMs, such as group velocity and localization, \kst{the MPhCR} provides an exciting platform for a broad range of photonics applications, including sensing/metrology, nonlinear optics, and cavity quantum electrodynamics.
\end{abstract}  

\maketitle
\noindent Chip-integrated optical micro-/nano-cavities have enabled numerous breakthroughs across the optical sciences~\cite{Vahala_Nature_2003}. These devices spatio-temporally enhance light-matter interactions in platforms suitable for integration and deployment, and have been established as foundational elements in quantum optics~\cite{Obrien_NatPhoton_2009}, nonlinear photonics~\cite{Strekalov_JOpt_2016}, optomechanics~\cite{Aspelmeyer_RevModPhys_2014}, and sensing~\cite{Vollmer_Nanophotonics_2012}. The metric for spatial confinement is mode volume ($V$), and the metric related to temporal enhancement is optical quality factor ($Q$). While there have been numerous cavity geometries studied, many of them fall into two categories: whispering gallery mode (WGM) cavities~\cite{Matsko_JSTQE_2006} based on total internal reflection at the device periphery, and photonic crystal (PhC) cavities based on localizing defects in one-dimensional or two-dimensional photonic lattices~\cite{Istrate_RevModPhys_2006}. In the former category, achieving many high-$Q$ WGMs across a wide wavelength range~\cite{Lu_NatPhoton_2019} is natural and requires no specific device geometry, as long as the device radius is sufficiently large and the device sidewall is smooth. In contrast, in PhC cavities, both multi-mode operation and high-$Q$ resonance(s) require careful and specific design of a device geometry, whose pattern must be preserved with certain accuracy in nanofabrication. Another advantage of WGM cavities is the relative ease of waveguide coupling through evanescent interaction. These two advantages have established WGM cavities as a major platform for broadband nonlinear optics, including Kerr frequency combs, optical frequency conversion, and other nonlinear wave mixing effects~\cite{Strekalov_JOpt_2016}. However, WGM cavities are unable to match the stronger spatial confinement provided by PhC defect cavities, whose $V$s are typically 10$\times$ to 100$\times$ smaller~\cite{Istrate_RevModPhys_2006}. PhC defect cavities can have sub-cubic-wavelength $V$ and are particularly suitable for single mode applications in cavity quantum electrodynamics~\cite{Obrien_NatPhoton_2009} and cavity optomechanics~\cite{Aspelmeyer_RevModPhys_2014}. \\
\indent Given the individual strengths of WGM and PhC defect cavities, it is desirable to combine these two geometries into one to utilize the best aspects of both platforms. Indeed, this aspiration has been pursued in various cavity geometries (See Extended Data Table~\ref{TabES1}). For example, one investigated geometry creates polygonal `disk' or `ring' shape line defects in two-dimensional PhCs, that is, photonic crystal `disk/ring' resonators (PCDRs/PCRRs)~\cite{Smith_APL_2001, Kim_APL_2002, Zhang_OL_2014}. Further pioneering work in conventional ring resonators named photonic crystal rings (PhCRs) has been conducted~\cite{Lee_OL_2012}, where etched air holes are introduced into a microring, with an emphasis on the generation of slow light modes~\cite{Lee_OL_2012, Zhang_PTL_2015, Gao_SciRep_2016, KML_OE_2017, Lo_OL_2018}. In such PhCRs, although the group velocity of light has been slowed down to enhance light-matter interaction, $Q$s are significantly degraded compared to state-of-the-art microrings, which offsets the benefits of using slow light in the first place.

Other work has incorporated small-amplitude gratings to microring sidewalls to control the frequencies of selected cavity modes~\cite{Lu_APL_2014} \xl{and has been used for single mode lasing~\cite{Arbabi_OE_15} and spontaneous pulse formation~\cite{Yu_NatPhoton_2021}, but without considering PhC-induced slow-light or localization effects.} \xl{Similar grating concepts have also been applied to whispering gallery resonators in many different contexts, including the generation of orbital angular momentum beams~\cite{Cai_Science_2012} and the exploration of parity-time concepts in microlasers~\cite{Feng_Science_2014}, but in general such works have not focused on the achievement of high-Q or strong mode localization through defect incorporation.}\\
\begin{figure*}[t!]
\centering\includegraphics[width=0.87\linewidth]{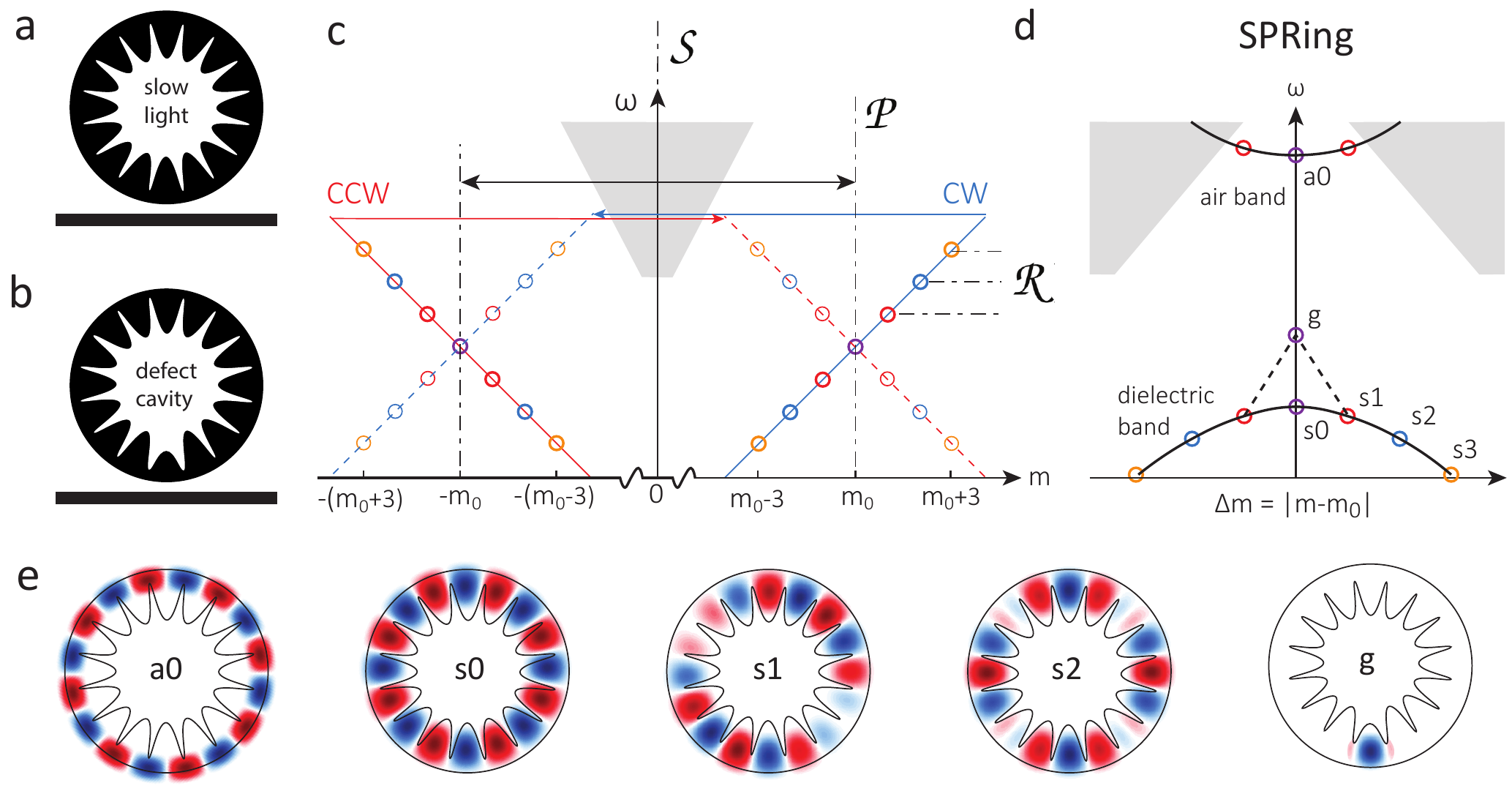}
\caption{\textbf{\xlt{MPhCR}: High-Q slow light photonic crystal microring.} \textbf{a-b,} Schematics of \kst{the MPhCR} for slow light (a) and photonic crystal defect (b) cavities. \textbf{c,} Three symmetries in the \xlt{MPhCR}: mirror symmetry ($\mathcal{S}$) of clockwise (CW) and counter-clockwise (CCW) propagating modes; rotational symmetry ($\mathcal{R}$) that allows modes with discrete momentum ($m$) and energy ($\omega$) only, and the periodic translation symmetry in the azimuthal direction from the PhC structure ($\mathcal{P}$), which couples CW and CCW modes with specific azimuthal momentum (i.e., $\pm m_0$). \textbf{d,} The band diagram created from these three symmetries. The air band is pushed towards the light cone (grey) and the dielectric band is well preserved. The modes in slow light devices are connected by solid lines, where $s0$ and $a0$ are band-edge modes. In dPhC devices, $s0$ shifts to $g$, as indicated by the dashed line, while other modes stay fixed in frequency. \textbf{e,} Qualitative illustrations of the dominant electric field component (e.g., $E_r$ for transverse-electric modes) of the labeled modes, with red and blue showing contrary phases and the darkness proportional to amplitude. Mode profiles from full numerical electromagnetic simulations are shown in Extended Data.}
\label{Fig1}
\end{figure*}
\indent In this work, we weave together aspects of WGM and PhC defect cavities in a novel way that retains the ease of realizing high-$Q$ and straightforward waveguide coupling of the WGM cavities while maintaining the ability to strongly manipulate the propagation and confinement of light associated with the PhC defect cavities. Our device, \kst{a `microgear' photonic crystal ring (MPhCR),} is a microring cavity in which a judiciously chosen periodic modulation to its inside boundary \kst{(resembling the shape of an internal gear), opening} a large bandgap at a targeted location in angular-momentum space. Within this platform we show slow-light modes with a high-$Q$ of $(1.1\pm0.1)\times10^6$ and a \ks{group velocity slowdown ratio ($SR$)~\cite{Krauss_JPhysD_2007,Lee_OL_2012} $\approx 10$}. We further show PhC defect modes based on localization of the slow-light mode at the band-edge. These localized defect modes have similarly high-$Q$s up to (5.6$\pm$0.1)$\times10^5$ and support intuitive frequency engineering and waveguide coupling like conventional WGMs. By marrying WGM and PhC concepts in a way that retains their respective advantages, \kst{the MPhCR} is a breakthrough platform for microcavity physics and applications. 

\medskip
\noindent \textbf{Introducing the \xlt{MPhCR} Design}
\indent Schematic illustrations of \kst{the MPhCR} are shown in Fig.~\ref{Fig1}(a,b), where (a) shows a slow light device and (b) shows a PhC defect (hereafter dPhC) device. Both devices have circular outside boundaries that are the same as a traditional microring that supports high-$Q$ WGMs with discrete angular momentum (described by azimuthal mode number $m$) and frequency ($\omega$). The slow light device has its inside boundary periodically modulated with a large amplitude, as shown in Fig.~\ref{Fig1}(a), which creates wide photonic band-gaps that support slow-light modes at and near the band-edges. The dPhC device is based on the slow light device design, with a localized defect incorporated in the periodic modulation (defect at center here, next to the couplling waveguide), as shown in Fig.~\ref{Fig1}(b).\\
\indent A \xlt{MPhCR} device has three types of symmetry, as shown in Fig.~\ref{Fig1}(c). $\mathcal{S}$ stands for mirror symmetry, that is, the symmetry of clockwise (CW) and counterclockwise (CCW) propagating light. $\mathcal{P}$ represents the angular momentum shift created by the periodic (in the azimuthal direction) PhC modulation, which is a transition of modes from CCW to CW (red arrow), or from CW to CCW (blue arrow). $\mathcal{R}$ represents the rotational symmetry due to the circular boundary condition, which leads to a quantized/integer angular momentum ($m$) and a discrete resonance frequency ($\omega_\text{m}$) for the WGMs.\\ 
\begin{figure*}[t!]
\centering\includegraphics[width=0.87\linewidth]{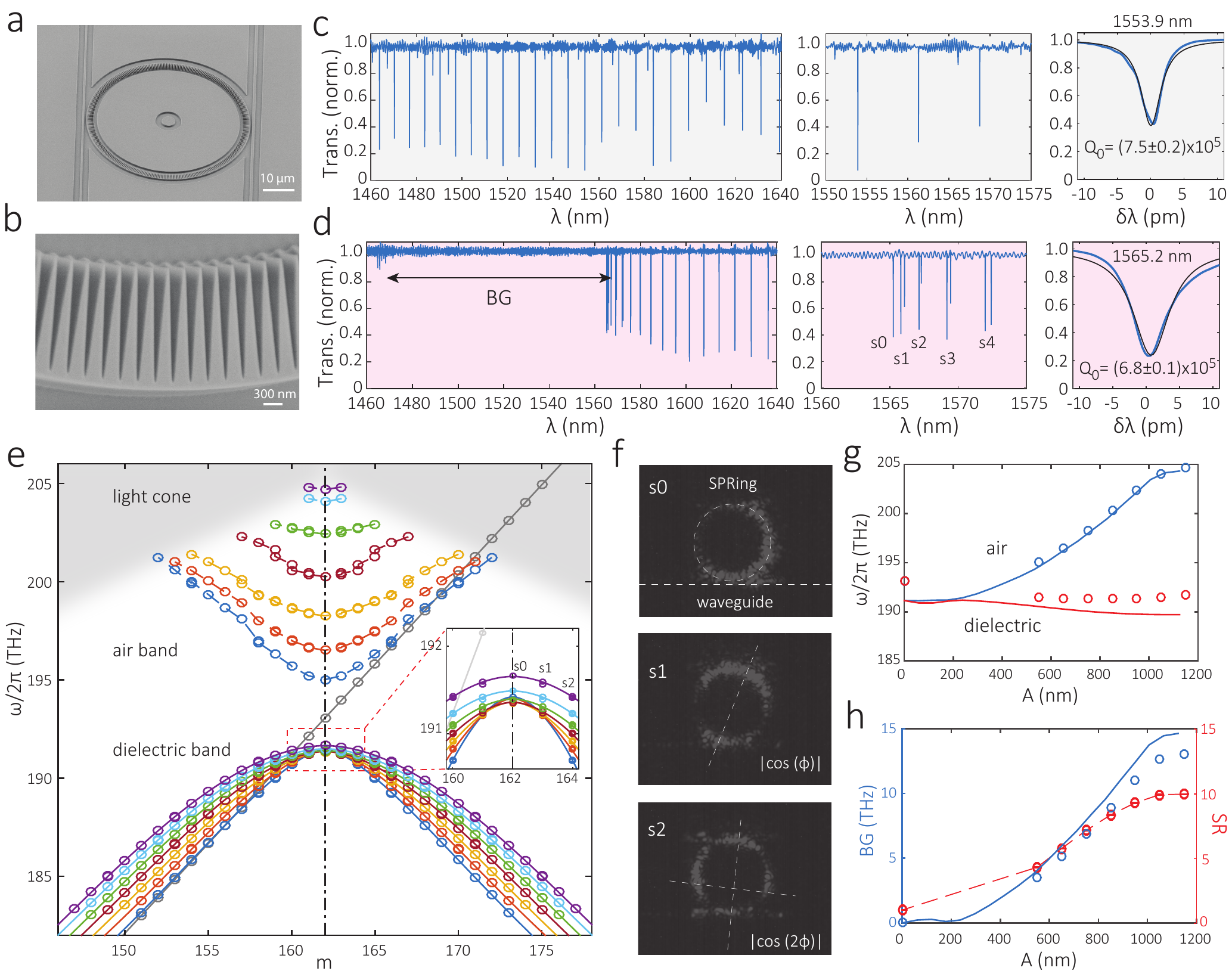}
\caption{\textbf{Slow light in \kst{the MPhCR}.} \textbf{a-b,} SEM image of a \xlt{MPhCR} with integrated waveguides, and zoom-in image of the PhC structure. \textbf{c-d,} Linear transmission spectrum of a control microring (no modulation) and a slow light \xlt{MPhCR} with modulation amplitude of $A$~=~1150~nm, respectively. The middle panels show a portion of the spectra near the $m$~=~162 mode, and the right panels show a zoom-in of the shortest wavelength mode from the middle panel (e.g., the $s0$ mode in (d)), with a nonlinear least squares fit to the data shown in black. The uncertainty in $Q_0$ is a one standard deviation value from this fit. \textbf{e,} Measured band diagrams of the control device (grey circles) and slow light \xlt{MPhCR} devices with $A$ from 550~nm (blue circles) to 1150~nm (purple circles), with a step of 100~nm. Solid lines are from fitting of the dielectric bands to a hyperbolic model (see Methods). Dashed lines are for guidance in viewing the air bands.  \textbf{f,} Infrared images of the light scattered from the slow light resonances $\{s_{0},s_{1},s_{2}\}$ in (d). The dashed lines in the first image outline the microring and coupling waveguide in use. The dashed lines in the other two images mark the antinodes (local maxima) of the scattered light intensity. s0 shows a more distributed pattern than s1 and s2 around the microring. We note that s0 is not as uniform as theoretically predicted, which might be due to weak defect localization but requires further study. In comparison, s1 and s2 show the expected two and four antinodes patterns, respectively, in the scattered field. \textbf{g,} The measured (circles) and simulated (lines) frequencies of the air band-edge and dielectric band-edge modes as a function of $A$. \textbf{h,} The bandgaps (blue) and slowdown ratios (red) as a function of $A$. The dashed line is for guidance in viewing $SR$.}
\label{Fig2}
\end{figure*}
\indent Applying a large inner sidewall modulation as shown in Fig.~\ref{Fig1}(a) results in a large photonic bandgap opening at $m~=~\pm m_\text{0}$ in Fig.~\ref{Fig1}(d), where the PhC modulation period is $\pi R$/$m_\text{0}$ (i.e., 2$m_\text{0}$ periods fit within the ring circumference). In \kst{the MPhCR}, two propagating modes ($m~=~\pm m_\text{0}$) are renormalized to two standing wave modes, labeled $a0$ (the air band-edge mode) and $s0$ (the dielectric band-edge mode), respectively. On either side of the band-edge, four modes with $m~=~\pm (m_\text{0} \pm 1)$ are hybridized to four new modes, two degenerate in the air band ($a1$), and two degenerate in the dielectric band ($s1$), where $s1$ can be viewed as having the same spatial profile as $s0$ with an additional cos($\phi$) modulation applied ($\phi$ is the azimuthal angle). The nature of $s2$ is similar, except the mode profile is modulated by an additional cos($2\phi$) with respect to $s0$. We note that the frequency splitting induced by the PhC modulation in a \xlt{MPhCR} results in high-frequency air band modes that begin to impinge on the light cone (grey). Starting with a slow light device, a dPhC device can be made by local perturbation of the modulation amplitude. The dPhC mode ($g$) is a localization of $s0$ mode and has a higher resonance frequency depending on the detailed parameters of the defect, as indicated by the dashed line in Fig.~\ref{Fig1}(d), while all other modes remain the same. Representative profiles for these optical modes are illustrated in a qualitative fashion in Fig.~\ref{Fig1}(e).  For illustration purposes, here we show an azimuthal mode number $m$~=~8, which is $\approx$ 20 times smaller than what we use in real devices. Full numerical simulations of the modes of the real geometries are shown in Methods and Extended Data Figs.~\ref{ExtFig1}-\ref{ExtFig2}. \\
\indent Scanning electron microscopy (SEM) images of fabricated \xlt{MPhCR}s in stoichiometric silicon nitride (Si$_3$N$_4$) are shown in Fig.~\ref{Fig2}(a,b), where Fig.~\ref{Fig2}(a) shows a ring radius of 25~$\mu$m, a thickness of 500~nm, a nominal average ring width of 1250~nm, and a modulation amplitude of $A$~=~1150~nm. A zoom-in SEM image of a slow light \xlt{MPhCR} shows the details of the structure in Fig.~\ref{Fig2}(b), including the large $A$ that leaves the ring width at only 100~nm in the narrowest part, similar in shape to `alligator' photonic crystals waveguides~\cite{Yu_APL_2014} previously studied. The device is fully etched through the Si$_3$N$_4$ layer with a smooth modulation profile and sidewall. The modulation period (length of one cell) is $\approx$ 485~nm, corresponding to 162$\times$2 cells in the circumference. See Methods for fabrication details.

\medskip
\noindent \textbf{Slow light in \kst{the MPhCR}}
Light-matter interactions can be enhanced by increasing the time over which the interaction occurs. Typically this is done by increasing the photon storage time (i.e., improving $Q$), for example, by using optical microcavities~\cite{Vahala_Nature_2003}. Another approach is by decreasing the group velocity of light, that is, using slow light effects, for example, based on  waveguide modes near photonic band-edges~\cite{Krauss_JPhysD_2007, Baba_NatPhoton_2008, Arcari_PRL_2014}. While each approach has been extensively studied individually, there should be situations where they can work together, that is, slowing down the group velocity of light while maintaining high $Q$. \xl{We emphasize that, in the photonic crystal microring, the slowdown factor should be included in the $Q$ already. In other words, the slow light devices might have higher optical $Q$s than the control device and provide additional benefits in applications. While there is evidence supporting the enhancement of $Q$ by slow light in the $Q < 5,000$ regime~\cite{Fujita_APL_2002}, all these} works have $Q$ values over two order of magnitude below the 10$^6$ values that are often achieved in conventional high-$Q$ microcavities without using slow light, which offsets the benefits of using slow light effects. For example, $Q$~=~2,000 and a slowdown ratio ($SR$) of 4 were achieved in pioneering work on photonic crystal microrings~\cite{Lee_OL_2012}. More recently, improved performance was reported~\cite{Gao_SciRep_2016} with $Q$~=~12,100 and $SR$~=~8. See Extended Data Table~\ref{TabES1} for details.\\
\begin{figure*}[t!]
\centering\includegraphics[width=0.87\linewidth]{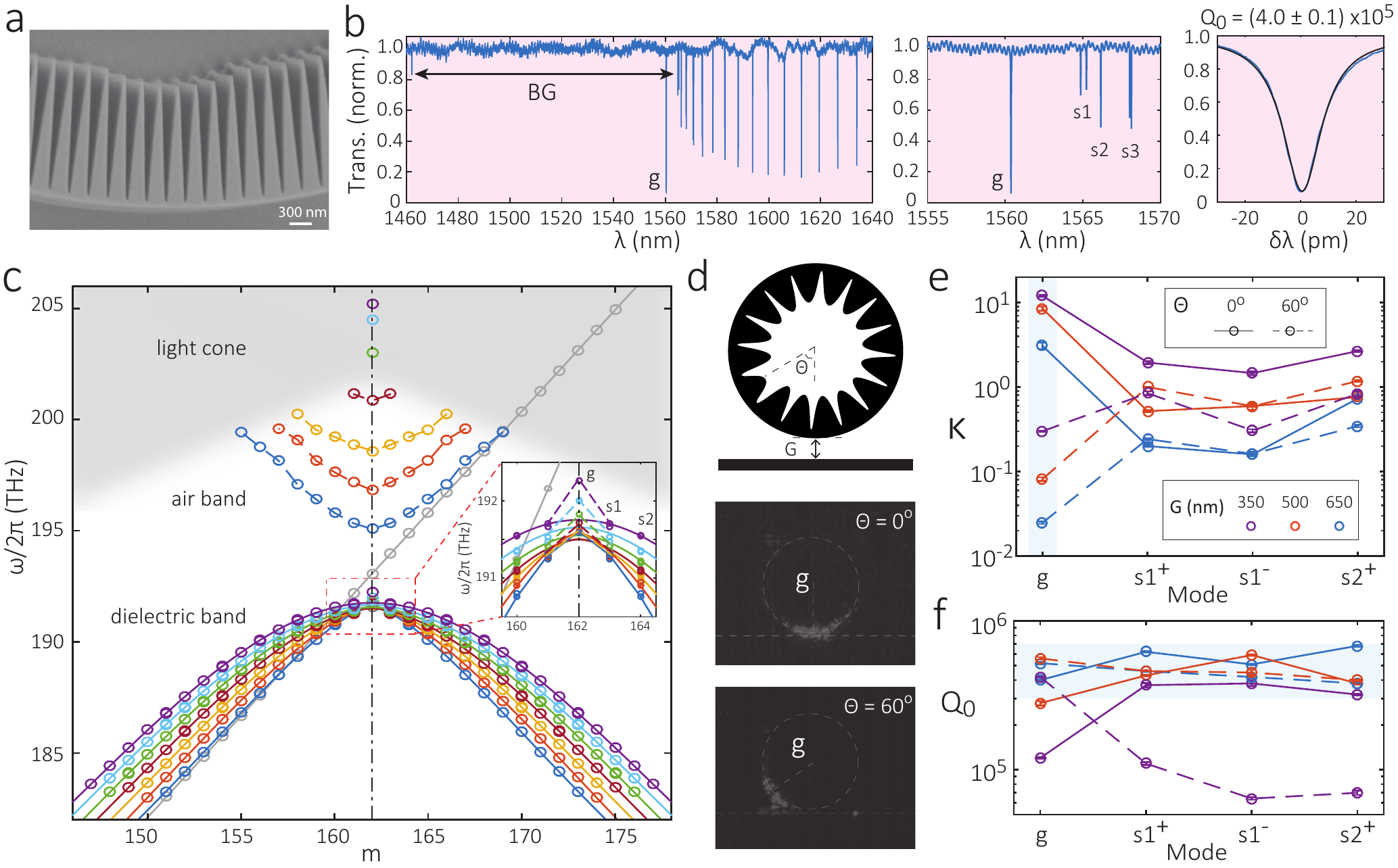}
\caption{\textbf{\kst{Defect modes in the MPhCR}.} \textbf{a,} Zoom-in SEM image of the \xlt{MPhCR} structure with a defect, that is, a quadratic grade of $A$ induced over $N$~=~8 cells. At defect center, the modulation depth of $A$ is $\Delta$~=~10~\%. \textbf{b.} Characterization of a defect \xlt{MPhCR} device incorporating a defect into the slow light design shown in Fig.~\ref{Fig2}(c), with $N$~=~48 and $\Delta$~=~10~\%. \textbf{c,} The band diagrams of defect \xlt{MPhCR} devices for the aforementioned $\Delta$ and $N$, with $A$ varying from 550~nm (blue) to 1150~nm (purple). The color scheme follows that of the slow light devices in Fig.~\ref{Fig2}(e). Inset is a zoom-in of the dielectric band-edges, and clearly shows the $g$ mode is separated from the dielectric band shifted into the bandgap. The solid lines for the dielectric bands are from the `hyperbolic' fitting (see Extended Data). The dashed lines are guidance for viewing. \textbf{d,} Schematic image (top) of a dPhC with its defect center angle ($\theta$) and the waveguide-ring coupling gap ($G$), and infrared images of the localized dPhC ground state ($g$) with defects centered (middle) and rotated (bottom), respectively. \textbf{e-f,} Coupling ratio ($K~=~Q_\text{0}/Q_\text{c}$) and intrinsic optical quality ($Q_\text{0}$) of the dPhC modes ($g$) and three slow light modes ($\{$$s1^{\pm}$, $s2^+$$\}$ in 6 \xlt{MPhCR} devices, that is, rotated (solid lines) and centered (dashed lines) 
defects with three $G$s.}
\label{Fig3}
\end{figure*}
\indent Our first objective in developing \kst{the MPhCR} is to achieve slow-light modes with high-$Q$s comparable to WGMs. In comparison to previous geometries with air holes within microrings~\cite{Lee_OL_2012,Gao_SciRep_2016}, \kst{the MPhCR} seems favorable for reducing scattering loss. We confirm this through spectral measurements presented in Fig.~\ref{Fig2}, where $Q$ and $SR$ are extracted from linewidths and free spectral ranges ($FSR$s), respectively (see Methods). In particular, as the $FSR$ is inversely proportional to the resonator mode group index, the reduction in $FSR$ relative to conventional microring modes is indicative of a corresponding increase in group index (and hence a decrease in group velocity). As in earlier works~\cite{Lee_OL_2012}, we use $SR$ as a figure of merit for both traveling wave and standing wave modes (both of which are commonly observed in conventional high-$Q$ WGMs without slow light effects, i.e., $SR$~=~1). For the latter, we are essentially considering the reduction in group velocity of the constituent counter-propagating traveling waves that make up the standing wave.

The conventional microring without modulation ($A$~=~0~nm) exhibits WGMs (mostly singlet resonances) with nearly-uniform $FSR$s ($\approx$ 0.91 THz or 7.4~nm), as shown in Fig.~\ref{Fig2}(c). The intrinsic optical quality factor ($Q_0$) is (7.5$\pm$0.2)$\times$10$^5$ (the right panel) for the mode at 1553.9~nm, whose azimuthal mode number ($m$~=~162) is targeted for inner sidewall modulation (the uncertainty of $Q_\text{0}$ comes from a nonlinear least squares fit of the transmission resonance; see Methods). When the modulation is very large, with $A$~=~1150~nm, as shown in Fig.~\ref{Fig2}(d), most air band modes are pushed into the light cone, leaving only three modes with low $Q$ and poor waveguide coupling. The dielectric band is well-preserved with a compressed spectrum with a higher spectral density of modes near the band-edge. This device has $Q_\text{0}$~=~(6.8$\pm0.1)\times$ 10$^5$ for the $s0$ mode. Infrared images of scattered light from $\{s$0, $s$1, $s$2$\}$ modes of Fig.~\ref{Fig2}(d) are shown in Fig.~\ref{Fig2}(f), and have azimuthal patterns that match those illustrated in Fig.~\ref{Fig1}(e).\\
\indent We construct the band diagrams of eight devices (including the two discussed) in Fig.~\ref{Fig2}(e). When $A$ increases, the air bands are pushed towards the light cone, but the dielectric band-edge stays fixed within 1 THz. The measured band-edge frequencies agree well with numerical simulations, as shown in Fig.~\ref{Fig2}(f) (the full simulated band structure for these devices is shown in Extended Data Fig.~\ref{ExtFig1}). Figure~\ref{Fig2}(h) shows the measured bandgaps, $BG=\omega(a0)-\omega(s0)$, and $SR$s for the $s0$ modes as a function of $A$. $SR$ increases with $BG$, and the device with the largest $BG$ exhibits $SR$~=~9.94~$\pm$~0.02 (the uncertainty of $SR$ comes from estimating free spectral ranges with split modes, see Methods) with the aforementioned $Q_\text{0}$~=~(6.8$\pm$0.1)$\times$ 10$^5$ for the $s$0 mode, which is $\approx$~50 times higher than previous works~\cite{Lee_OL_2012, KML_OE_2014, KML_OE_2017,Gao_SciRep_2016}. Extended Data Fig.~\ref{ExtFig3}(b) shows $SR$ for all values of $A$ and all dielectric band modes near the band-edge, indicating how $s1$ and $s2$ also exhibit significant, but smaller $SR$ values.

An important metric for slow light applications (e.g., interferometry~\cite{Shi_JOSAB_2008}) is $n_\text{g}/\alpha$, where $n_\text{g}$ is the group index and $\alpha$ is the loss coefficient (directly proportional to $Q_\text{0}$; see Methods). The $s0$ mode in the $A$=1150~nm device has $n_\text{g}/\alpha$~=~(17$\pm$1)~cm (the uncertainty is propagated from that of $Q_\text{0}$), which is a significant improvement in comparison to earlier works. Moreover, the optical $Q$ can be further improved. For example, in Extended Data Fig.~\ref{ExtFig4}, we show a slow-light device with high-$Q$ s0 mode of $Q_\text{0} = (1.1\pm0.1)\times10^6$, with $SR~\approx$~10 and $n_\text{g}/\alpha~\approx$~28. Such high-$Q$, or equivalently $n_\text{g}/\alpha$, represents $>~50\times$ enhancement of previous record; see Extended Data Table~\ref{TabES1} for comparison. \kst{The MPhCR} is thus a promising platform for further investigation of slow light applications, including interferometry, telecommunications, and sensing~\cite{Shi_JOSAB_2008, Shi_PRL_2007, Boyd_JMO_2009, Vollmer_Nanophotonics_2012}.

\medskip
\noindent \textbf{\kst{Defect modes in the MPhCR}}
\xl{While $n_\text{g}/\alpha$ is an important figure-of-merit for slow light application as discussed in the previous section, $Q/V$ is the metric that is critical} for cavity-enhanced light-matter interactions, including nonlinear and quantum optical processes (for example, Purcell enhancement). In a microring, this value can be optimized by reducing the radius of the microring while maintaining relatively low radiation/scattering loss, with bending loss ultimately limiting the smallest $V$s achievable. In this section, we show how a localized PhC defect (dPhC) can be introduced into the slow light devices to further reduce the mode volume ($V$) and improve $Q/V$ by $\approx$ 10 times with respect to our conventional microrings. For example, Fig.~\ref{Fig3}(a) is a zoom-in SEM image that shows how to derive a dPhC device from a slow light device  such as that in Fig.~\ref{Fig2}(b). Here, a defect region is introduced in which the modulation amplitude is varied quadratically across eight cells (four on either side), with a maximum modulation depth deviation ($\Delta$) of 10~\% of $A$ at the defect center. This intuitive defect design induces a bound simple harmonic potential well that supports a ground state (g) dPhC mode. Figure~\ref{Fig3}(b) shows the transmission of a dPhC mode ($g$) created similarly to Fig.~\ref{Fig3}(a) but across 48 cells, with $Q_\text{0}~=~(4.0\pm0.1)\times10^5$ that is similar to that of the slow-light modes. The highest $Q_\text{0}$ observed is (5.6$\pm$0.1)$\times10^5$; see Extended Data Fig.~\ref{ExtFig5}(e). The creation of high-$Q$ dPhC modes through perturbation of the slow-light cavities is possible without specific detailed design, a heritage from the WGM side that stands in contrast to most dPhC cavities, where detailed numerical simulations are needed to optimize designs to support high-$Q$~\cite{Vuckovic_PRE_2001,Srinivasan_OE_2002,Englund_OE_2005,Asano_JSTQE_2006,Quan_OE_2011}. By introducing defects to all eight slow-light designs in Fig.~\ref{Fig2}(e) in the aforementioned fashion, the defect band diagram in Fig.~\ref{Fig3}(c) is generated. A zoom-in on the dielectric band-edge (Fig.~\ref{Fig3}(d)) shows how the dominant change with respect to the band diagram for the unperturbed slow-light devices (Fig.~\ref{Fig2}(e)) is a shift of the $s0$ mode into the band-gap to create $g$, while the neighboring slow-light modes (e.g., $\{s1,s2,s3\}$) are largely unperturbed. In Extended Data Fig.~\ref{ExtFig5}(a-c), we show the frequency control of $g$ by parameters other than $A$, including $m$, $N$, and $\Delta$.\\
\indent We have confirmed in simulation the existence of these experimentally observed $g$ modes, and also calculated their mode volumes. Extended Data Fig.~\ref{ExtFig2} presents full numerical simulations of the dPhC mode profiles and resulting $V$s, using the cavity QED definition for $V$ based on electric dipole coupling~\cite{Vuckovic_PRE_2001} (see Methods). $V=3.3(\lambda/n)^3$ to 5.5$(\lambda/n)^3$ in dPhC modes, a reduction $>10\times$ relative to the slow-light band-edge modes, and as expected, the amount of localization is dependent on the specific characteristics of the potential. \ks{We anticipate that numerical optimization, and incorporation of basic methods to increase modal confinement, such as a lower air cladding or slot-mode architectures, can further reduce $V$ in this platform, while $\sim(\lambda/n)^3$ values are likely possible in higher index platforms such as silicon-on-insulator.}\\
\indent Applications exploiting enhanced light-matter interactions in optical microcavities also require effective waveguide input/output coupling. We next consider the ability to couple to dPhC modes in a controlled fashion, an attribute passed to \kst{the MPhCR} from its WGM heritage. In Fig.~\ref{Fig3}(d), we illustrate two parameters that control the evanescent coupling between a \xlt{MPhCR} and a waveguide, namely the defect angle ($\theta$) and the waveguide-ring gap ($G$). We also show two infrared images of the scattered light from a defect-rotated ($\theta$~=~60$^o$) and defect-centered ($\theta$~=~0$^o$) device. The measured coupling ratio $K~=~Q_\text{0}/Q_\text{c}$, for both of the dPhC mode ($g$) and the slow-light modes $\{s_{1}^{\pm},s_{2}^{+}\}$, is plotted in Fig.~\ref{Fig3}(e) for six cases -- three $G$s and two $\theta$s. It is clear that the coupling to $g$ is very different than that to $\{s_{1}^{\pm},s_{2}^{+}\}$, because of its localization. In the rotated case, $g$ has a weaker coupling ($\approx$ 1/10$\times$ in $K$) than $\{s_{1}^{\pm},s_{2}^{+}\}$. In contrast, in the centered case, $g$ has a stronger coupling than $\{s_{1}^{\pm},s_{2}^{+}\}$ ($\approx$ 10$\times$ in $K$). For a \xlt{MPhCR} with $G$~=~350~nm, $Q_\text{c}$ can reach $10^4$ (see Extended Data Fig.~\ref{ExtFig5}), and even stronger coupling could be realized by using smaller $G$, optimized waveguide width, and pulley coupling~\cite{Li_NatPhoton_2016, Greg_OL_2019}.\\
\indent As shown in Fig.~\ref{Fig3}(e), $Q_\text{0}$ values for dPhC modes are similar to the slow light modes in the under-coupled cases, and lie in a range of $3 \times 10^5$ to $8 \times 10^5$ (shaded area). We note that these slow light modes and dPhC modes have $Q$s similar to WGMs in the control devices (within a factor of 2), illustrating how in \kst{the MPhCR}, slow-light effects and strong modal confinement can be effectively realized in a WGM platform. 
\indent \xl{Finally, we note that the dPhC mode on its own does not require a full microring to exist, unlike the slow light modes and conventional WGMs. However, retaining the full microring structure enables the co-existence of the dPhC mode with these other modes, which will be essential for nonlinear optics and other multi-mode applications. We also note that the continuous transition of the dPhC mode to the band-edge slow light modes seems to be related to bound-state-in-continuum phenomena.}

\medskip
\noindent \textbf{Discussion} We have demonstrated a new microcavity platform for combining some of the most important aspects of WGM and PhC cavities. In essence, our \xlt{MPhCR} design controllably modifies fundamental characteristics of targeted microring WGMs, such as their group velocity and localization length scale, while maintaining their high-$Q$ and straightforward waveguide coupling. The design of \xlt{MPhCR} geometries is highly intuitive, with the devices we have shown not requiring any detailed numerical modeling to achieve high-$Q$, large $SR$, or strong mode localization (numerical modeling was used simply to verify device performance). Going forward, the coexistence of slow light, dPhC and conventional WGM modes is promising for nonlinear optics, for example, optical parametric oscillation~\cite{Lu_Optica_2019, Marty_NatPhoton_2021}; moreover, incorporating multiple photonic bandgap frequencies within a single device is feasible, similar to the ability to selectively introduce backscattering for multiple WGMs~\cite{Lu_PRJ_2020}. Another effort could be in focusing on the performance of air band modes to, for example, develop devices for coupling to neutral atoms~\cite{Yu_APL_2014,Douglas_NatPhoton_2015}. \xl{As discussed in a recent review~\cite{ji_methods_2021}, while silicon nitride is generally considered as an especially high-$Q$ platform in integrated photonics, the $Q$ achieved is a strong function of the amount of modal confinement within the waveguiding layer, which is determined by the waveguide cross-section size and the cladding layer refractive indices. In a moderate-confinement silicon nitride microring, state-of-the-art $Q$s are in the 10$^6$ range; in high-confinement and low-confinement silicon nitride microrings, state-of-the-art $Q$s are much higher, in the 10$^7$ and 10$^8$ range, respectively. 
While we demonstrate \kst{the MPhCR} in a moderate-confinement silicon nitride microring in this work, it is of interest to implement \kst{the MPhCR} in high-confinement and low-confinement silicon nitride microrings for both higher $Q$s and also additional applications such as ultra-narrow-band filters, delay lines, and stimulated Brillouin scattering.} Finally, the development of \kst{the MPhCR} in other platforms, such as silicon carbide~\cite{Lukin_NatPhoton_2019} and gallium arsenide~\cite{lodahl_interfacing_2015} could lead to new opportunities to control and exploit strong interactions between confined optical fields and single quantum emitters. 

\bibliographystyle{osajnl}

\bigskip
\bigskip
\bigskip
\noindent \textbf{Methods}

\medskip
\noindent \textbf{Simulation of \xlt{MPhCR} devices} \\
\noindent The inner sidewall modulation is given by the function:
\begin{eqnarray}
R_\text{in}~=~R_\text{out}-RW+A(1-2|\text{cos}(m_\text{0} \cdot \phi)|),
\end{eqnarray}
where $R_\text{out}$~=~25~$\mu$m, $RW$~=~1250~nm, $A$ is the modulation amplitude ($A<RW$), $\phi$ is the azimuthal angle, and $m_\text{0}$~=~162 or 164 is the targeted azimuthal mode number to open the bandgap.

Simulated band diagrams in Extended Data Fig.~\ref{ExtFig1}(a) are obtained with MIT Photonic Bands (MPB)~\cite{Johnson_OE_2001}. 
The structure in simulation is a rectangular unit cell with equivalent sidewall contours of the \xlt{MPhCR}, with periodic boundary condition imposed~\cite{KML_OE_2014}. Neglecting the bending effect in this simulation is an acceptable approximation because the microrings we use have a sufficiently large radius. All simulations use refractive indices of 1.98 and 1.44 for the Si$_3$N$_4$ core and the SiO$_2$ substrate, respectively. Simulated band-edges/gaps in Fig.~\ref{Fig2}(g,h) are extracted from Extended Data Fig.~\ref{ExtFig1}(a).

To consider the microring bending effect and estimate the effective mode volume of the dPhC modes, we set up a 3D finite-element-method (FEM) simulation. For example, as shown in Extended Data Fig.~\ref{ExtFig1}(b,c) we simulate a unit cell with boundary conditions of perfect magnetic/electric conductors for the azimuthal-cutting planes ($\phi~=~\phi_\text{0}\pm\pi/(2m)$) for $s$0/$a$0 modes, respectively, and scattering conditions for other boundaries. In Extended Data Fig.~\ref{ExtFig2}(c-e), we simulate half of a photonic crystal defect structure, with the boundary condition of the plane $\phi~=~\phi_\text{0}$ containing the defect center to be perfect magnetic conductor for the $g$ mode (localization of $s$0), and scattering conditions for the other boundaries. The mode profiles are unfolded in Extended Data Fig.~\ref{ExtFig2}(d,e) for viewing. The mode volumes in Extended Data Fig.~\ref{ExtFig1} and \ref{ExtFig2} are calculated according to the definition used in cavity QED in the context of coupling between the cavity mode and an embedded electric dipole, and is given by:
\begin{eqnarray}
V~=~\frac{\iiint{\vec{E} \cdot \vec{D}}dv}{max(\vec{E} \cdot \vec{D})}.
\end{eqnarray}
where $\vec{E}$ and $\vec{D}$ are the electric field and electric displacement field, respectively, and the volume integral is taken over the whole space. The confinement factors are estimated by the energy integrated in the Si$_3$N$_4$ core divided by that over the whole space: 
\begin{eqnarray}
\eta~=~\frac{\iiint_{core}{\vec{E} \cdot \vec{D}}dv}{\iiint{\vec{E} \cdot \vec{D}}dv}.
\end{eqnarray}

\medskip
\noindent \textbf{Device fabrication} \\
\noindent The device layout was implemented through use of the Nanolithography Toolbox, a free software package developed by the National Institute of Standards and Technology Center for Nanoscale Science and Technology~\cite{Balram_JResNIST_2016}. The ${\rm Si_3N_4}$ device layer was grown by low-pressure chemical vapor deposition on top of a nominal 3~$\mu$m thick ${\rm SiO_2}$ layer, grown via thermal wet oxidation of Si, on a 100~mm diameter Si wafer. The wavelength-dependent refractive index and layer thicknesses were measured using a spectroscopic ellipsometer, with the data fit to an extended Sellmeier model. The device patterns are created in positive-tone resist by a 100 keV electron-beam lithography system, and then transferred to ${\rm Si_3N_4}$ by reactive ion etching using a ${\rm CHF_3/O_2}$ chemistry. The devices are then chemically cleaned in multiple steps to remove deposited polymer and remnant resist. An SiO$_2$ lift-off process based on photolithography and plasma-enhanced chemical vapor deposition with an inductivel-coupled plasma source is performed so that the resonators have a top air cladding while the input/output edge-coupler waveguides have a top ${\rm SiO_2}$ cladding to create more symmetric modes for coupling to optical fibres. The facets of the chip are then polished for lensed-fibre coupling. After being polished, the chip is annealed again at $\approx$~1000~${\rm ^{\circ} C}$ in a ${\rm N_2}$ environment for 4 hours.

\medskip
\noindent \textbf{Quality factor estimation for \xlt{MPhCR} devices} \\
\noindent 
Whispering gallery modes in high-$Q$ microring cavities are typically travelling waves, with a transmission at the resonance center is given by:
\begin{eqnarray}
T~=~|\frac{1/Q_\text{0}-1/Q_\text{c}}{1/Q_\text{0}+1/Q_\text{c}}|^2~=~|\frac{1-K}{1+K}|^2.
\end{eqnarray}
where $K~=~Q_\text{0}/Q_\text{c}$ is the coupling ratio in Fig.~\ref{Fig3}(e). Therefore when $K~=~1$ (critically coupled), $T~=~0$; when $K \ll 1$ (strongly undercoupled) or $K \gg 1$ (strongly overcoupled), $T \approx 1$.

The slow light modes and dPhC modes in \xlt{MPhCR} devices are standing-wave modes instead of propagating modes, and due to their clockwise and counterclockwise components, couple to both the forward and backward direction of the waveguide. Therefore, the fitting of these modes needs to be modified with an additional coupling channel. Such side-coupled standing wave cavities have been considered in numerous contexts, including PhC defect cavities~\cite{Xu_PRE_2000,Afzal_JOSAB_2019} and WGM cavities in the limit of strong surface-roughness-induced backscattering~\cite{kippenberg_modal_2002}. The transmission at the resonance center is given by:
\begin{eqnarray}
T~=~|\frac{1/Q_\text{0}+1/Q_{\text{c}^-}-1/Q_{\text{c}^+}}{1/Q_\text{0}+1/Q_{\text{c}^-}+1/Q_\text{c}^+}|^2.
\end{eqnarray}
where $Q_{\text{c}^+}$ and $Q_{\text{c}^-}$ denote coupling quality factors to the forward and backward direction, respectively. When this coupling is symmetric, that is, $1/Q_\text{c}~=~2/Q_\text{c}^+~=~2/Q_\text{c}^-$, the equation is reduced to:
\begin{eqnarray}
T~=~|\frac{1/Q_\text{0}}{1/Q_\text{0}+1/Q_\text{c}}|^2~=~|\frac{1}{1+K}|^2.
\end{eqnarray}
In the above equation, we keep the same definition of $K$ as in the traveling wave case, which means that when $K \gg 1$, $T \approx 0$ for the standing-wave modes (including slow light modes and dPhC modes), in sharp contrast to $T \approx 1$ for traveling-wave modes (such as conventional whispering gallery modes) with single waveguide coupling. See Extended Data Fig.~\ref{ExtFig5} for experimental results.

\medskip
\noindent \textbf{`Hyperbolic' parameter ($\zeta$) in band diagram fits} \\
The calculated $SR$s are closely related to $\zeta$, the `hyperbolic' index in band diagram fitting, given by $\omega(m)~=~ \omega_\text{0}+\omega_{1}/[(m-m_\text{0})^{\zeta}+m_\text{1}]$, where $\omega(m)$ is the dielectric band frequency as a function of azimuthal mode number $m$, $m_\text{0}$ is a known parameter from the modulation pattern, and $\omega_\text{0}$, $\omega_\text{1}$, and $\zeta$ are parameters to fit. $\omega_\text{0}$ is the $m_\text{0}$ ``asymptotic" center frequency of the hyperbolic curve. $\omega_\text{1}$/$m_\text{1}$ is the frequency separation between the $s0$ mode ($m~=~m_\text{0}$) at the dielectric band-edge and $\omega_\text{0}$. $\zeta$ describes the curvature of the dielectric band, which is closely related to $SR$. Naturally, $\zeta$~=~1 when there is no (or a very small) bandgap opened, while $\zeta$ $\approx$ 2 is the apparent limit of our design, in keeping the dielectric band-edge fixed and pushing the air band-edges to the light cone and reducing its curvature to be flat ($\zeta$~=~0). 

\medskip
\noindent \textbf{Slowdown ratio ($SR$)} \\
\noindent The slowdown ratio $SR$ presented in the main text is calculated from the cavity free spectral ranges ($FSR$s)~\cite{Lee_OL_2012}, and is normalized to the $FSR$ that is furthest from the band-edge, as shown in Extended Data Fig.~\ref{ExtFig3}(a). Typically there are two standing-wave modes in each mode set, that is $s$1$^{\pm}$, $s$2$^{\pm}$, etc., whereas $s$0 is always a singlet standing-wave mode. The assignment of $SR>1$ for standing wave modes is essentially an estimate of the $SR$ for their constituent traveling waves. We use the average values of the frequencies of the split modes for calculation of $SR$. \xl{These average frequencies represent the spectral locations} of the constituent traveling wave modes prior to backscattering-induced modal coupling, \xl{which forms the split standing wave modes}. The mode splitting \xl{is therefore used to estimate} the uncertainty of $SR$ (the error bars in Fig.~\ref{Fig2}(h)). The $SR$ values are underestimated for $s0$, as they use the wider $FSR$ adjacent to the mode, which represents the $SR$ of a frequency in between that of $s$0 and $s$1 ($\Delta m$~=~0.5) instead of that of $s$0 ($\Delta m$~=~0). Therefore, the maximum measured $SR$~=~9.9 is a lower-bound estimate for what is achievable in \xlt{MPhCR}s, as a method to extract a more accurate value of $SR$ beyond what the $FSR$ analysis can yield requires further investigation.

\medskip
\noindent \textbf{Metric for high-$Q$ slow light ($n_\text{g}/\alpha$)} \\
\noindent As mentioned in the main text, an important metric for high-$Q$ slow light is $n_\text{g}/\alpha$, due to its relevance for practical applications of slow light. For example, in slow light interferometry, it is important not only that the light is propagating more slowly (i.e., smaller $c/n_\text{g}$, where $c$ is speed of light in vacuum), but also that the light can propagate through sufficient length, in other words, the loss per unit length ($\alpha$) needs to be small. Although our high-$Q$ slow light devices at present cannot be directly used in slow light interferometry, for a guidance of future study within the context of slow light applications, we give the following estimate of $n_\text{g}/\alpha$:
\begin{eqnarray}
\frac{n_\text{g}}{\alpha}~=~\frac{Q_\text{0}\lambda}{2\pi}.
\end{eqnarray}
In the device presented in Fig.~\ref{Fig2}(d), we have $n_\text{g}/\alpha$ = (0.17~$\pm$~0.01)~m, where the uncertainty is a one standard deviation value originating from the nonlinear least squares fit of the transmission data to extract $Q_{\text{0}}$. We can see that $n_\text{g}/\alpha$ is directly proportional to $Q_\text{0}$, and that our work seems to suggest a $\approx$10$\times$ increase of $\alpha$ at the band edge for current devices, as we have an increase of $n_\text{g}$ $\approx$10$\times$ with similar $Q_\text{0}$. Going forward, it will be important to investigate whether we can further decrease the loss $\alpha$ at the band edge, or our work is already approaching the theoretical limit of scattering for slow light~\cite{Hughes_PRL_2005}.

\medskip
\noindent \textbf{Data availability} The data that supports the plots within this paper and other findings of this study are available from the corresponding authors upon reasonable request.

\medskip
\noindent \textbf{Acknowledgements} This work is supported by the DARPA SAVaNT and NIST-on-a-chip programs. X.L. acknowledges support under the Cooperative Research Agreement between the University of Maryland and NIST-PML, Award no. 70NANB10H193. The authors thank Zhimin Shi and Vladimir Aksyuk for helpful discussions.

\medskip
\noindent \textbf{Author contributions} X.L. led the design, fabrication, and measurement of the \xlt{MPhCR} devices. A.M. led the simulation with the help from X.L., and all authors participated in analysis and discussion of results. X.L. and K.S. wrote the manuscript with assistance from A.M., and K.S. supervised the project.

\clearpage
\renewcommand{\figurename}{Extended Data Fig.}
\renewcommand{\tablename}{Extended Data Tab.}
\setcounter{figure}{0}

\begin{table*}[t!]
\caption{\label{tab:table1} Comparison to previous works in combing a PhC and a microring, referred to in the literature as photonic crystal ring (PhCRs) and photonic crystal disk/ring resonators (PCDRs/PCRRs), and whose structures are based on microrings and 2D PhCs, respectively. The original work in a PhCR and a PCRR are in bold. PhCRs typically support multiple modes in a WGM fashion, and therefore have $SR$s listed. PCRRs only have single or few modes and $SR$s are not analyzed. Although not listed here, PCRRs typically have much smaller mode volumes than PhCRs. We note that reference~\cite{KML_OE_2017} has $Q$ and $SR$ reported in different devices. Overall, we find that \kst{the MPhCR} has the highest $n_\text{g}/\alpha$, which is an important metric for slow light application (see Methods).}
\begin{ruledtabular}
\begin{tabular}{ccccccccc}
device & year & core material & geometry &  RR$\times$RW$\times$RH ($\mu$m) & $\lambda$ (nm) & $Q_\text{0}$ & $SR$ & $n_\text{g}/\alpha$ (cm)\\
\hline
\xlt{MPhCR} & 2021 & Si$_3$N$_4$ & modulated microring & 25$\times$(0.1-2.3)$\times$0.5 & 1556 & 1.1$\times10^6$ & 8-12 & 28\\
\hline
\textbf{PhCR~\cite{Lee_OL_2012}} & \textbf{2012} & \textbf{Si} & \textbf{holes in microring} & 7.2$\times$0.45$\times$0.22 & 1508 & $2\times10^3$ & 4-5 & 0.05\\
PhCR~\cite{Zhang_PTL_2015} & 2015 & Si & holdes in racetrack ring & $\approx$~6$\times$0.42$\times$0.22 & 1554 & $8.3\times10^3$ & 11 & 0.2\\
PhCR~\cite{Gao_SciRep_2016} & 2016 & Si & holes in microring & 20$\times$0.68$\times$0.22 & 1558  & $1.2\times10^4$ & 8 & 0.3\\
PhCR~\cite{KML_OE_2017} & 2017 & Si & holes in microring & 20$\times$0.45$\times$0.22 & 1550  & $2.3\times10^4$ & 7 & 0.6\\
PhCR~\cite{Lo_OL_2018} & 2018 & Si & holes in microring & 7.2$\times$0.45$\times$0.22 & 1520  & $>1.5\times10^3$ & 4 & $>0.04$\\
\hline
\textbf{PCDR~\cite{Smith_APL_2001}} & \textbf{2001} & \textbf{GaAs} & \textbf{hexagonal `disk' in 2D PhC} & $\approx~1.8\times1.8\times0.22$ & 1000 & $>1\times10^3$ & - & $>$~0.02\\
\textbf{PCRR~\cite{Kim_APL_2002}} & \textbf{2002} & \textbf{InGaAsP} & \textbf{hexagonal `ring' in 2D PhC} & $\approx$~4$\times$1$\times$0.2 & 1625  & $>2\times10^3$ & - & $>$~0.05\\
PCRR~\cite{Zhang_OL_2014} & 2014 & Si & hexagon `ring' in 2D PhC & $\approx$~3$\times$1$\times$0.22 & 1554 & $7.5\times10^4$ & - & 1.9
\end{tabular} ~\label{TabES1}
\end{ruledtabular}
\end{table*}

\begin{figure*}[t!]
\centering\includegraphics[width=1.00\linewidth]{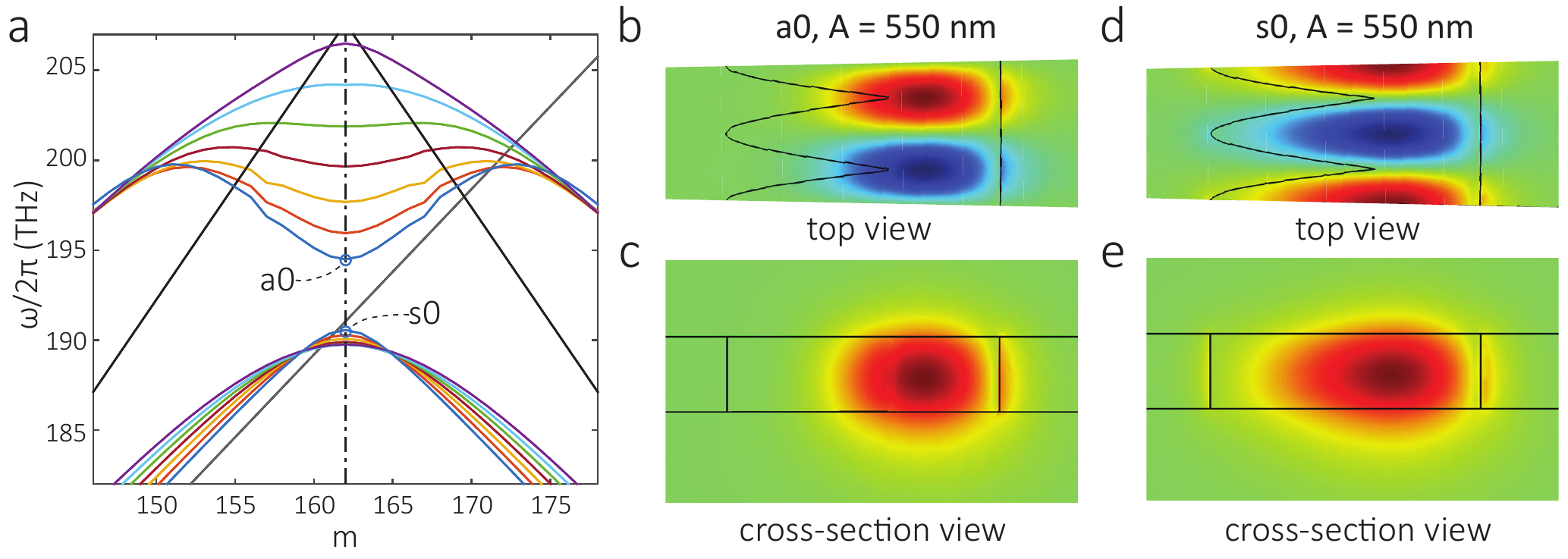}
\caption{\textbf{Simulated band diagrams and mode profiles of \xlt{MPhCR} cavities.} \textbf{a,} Simulated photonic bands for the \xlt{MPhCR} devices experimentally studied in Fig.~\ref{Fig2}(e), with $A$ varying from 550~nm (blue) to 1150~nm (purple), as well as the control device (grey). The solid black lines correspond to the light cone with n~=~1.44, the refractive index of the silica substrate. The slow light modes at the air band-edge ($a$0) and dielectric band-edge ($s$0) of the device with $A$~=~550~nm (blue) are highlighted by open circles. \textbf{b-e,} Finite-element method (FEM) simulations of a unit cell for the $a$0 and $s$0 modes, displayed from a top view (b,d) over one period of the optical field as well as in a cross-section view (c,e). Both modes have dominant electric fields in the radial (horizontal) direction. The resonance frequencies are 193.9~THz and 191.1~THz, within 1~THz of the experimental data (Fig.~\ref{Fig2}(e,g)). These $a$0 and $s$0 modes have mode volumes of $V$~=~0.29$(\lambda/n)^3$ and $V$~=~0.35$(\lambda/n)^3$ with confinement factors of $\eta$~=~71~\% and $\eta$~=~84~\%, respectively. The mode volumes for $a$0 and $s$0, containing 162 cells, are 47$(\lambda/n)^3$ and 56$(\lambda/n)^3$. See Methods for more details on the simulations and definition of mode volume and confinement factor.}
\label{ExtFig1}
\end{figure*}

\begin{figure*}[t!]
\centering\includegraphics[width=1.00\linewidth]{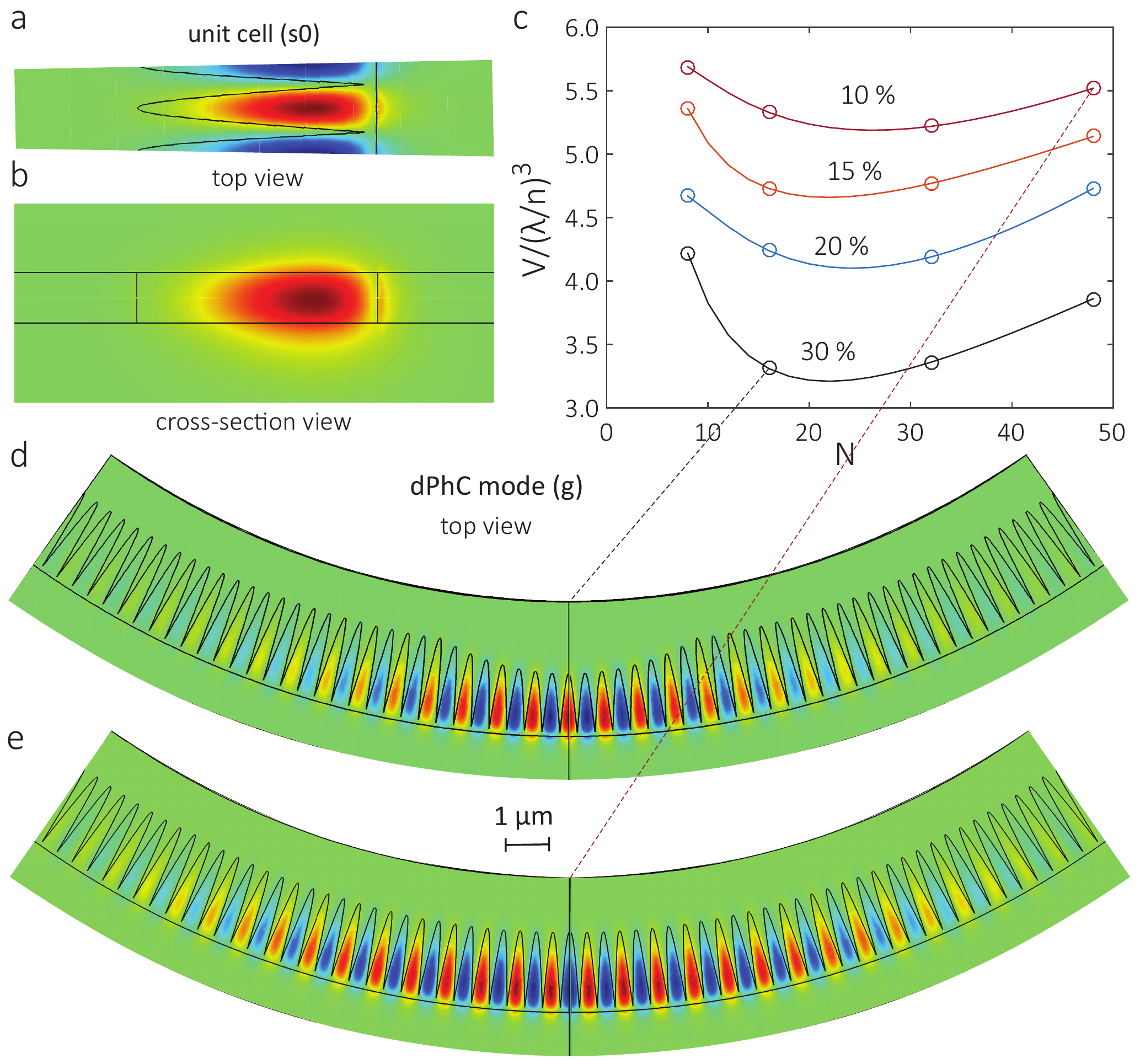}
\caption{\textbf{Simulated mode profiles and mode volumes for the unit cell and the photonic crystal defect modes built on this cell.} \textbf{a,b,} Top view and cross-section view of $s$0 FEM-simulated mode profiles in a unit cell with $m$~=~162 and $A$~=~1150~nm. The calculated resonance frequency is 191.0~THz, within 1~THz of the experimental data (Fig.~\ref{Fig2}(e)). The mode volume for a unit cell is 0.20~$\mu$m$^3$, that is, 0.40$(\lambda/n)^3$, with $\eta$~=~86~$\%$ confinement factor in Si$_3$N$_4$. The mode volume of $s$0 over the whole ring is 64$(\lambda/n)^3$. \textbf{c,} Photonic crystal (PhC) defects can be built based on (a). The volumes for such dPhC cavity modes are created by a reduction in $A$ ($\Delta$) by 10~$\%$ (magenta) to 30~$\%$ (black) at the defect center, and a quadratic grading in $A$ over $N$ cells is shown. The mode volumes can be further optimized by using a larger $\Delta$ and an optimized $N$, or using material stacks with larger refractive index contrasts, for example, silicon on insulator (SOI). \textbf{d,e,} Two examples of the dPhC mode ($g$) with $\{$$N$, $\Delta$$\}$~=~$\{$16, 30~$\%$$\}$ and $\{$48, 10~$\%$$\}$, with mode volumes of 3.3$(\lambda/n)^3$ and 5.5$(\lambda/n)^3$, respectively.} 
\label{ExtFig2}
\end{figure*}

\begin{figure*}[t!]
\centering\includegraphics[width=1.00\linewidth]{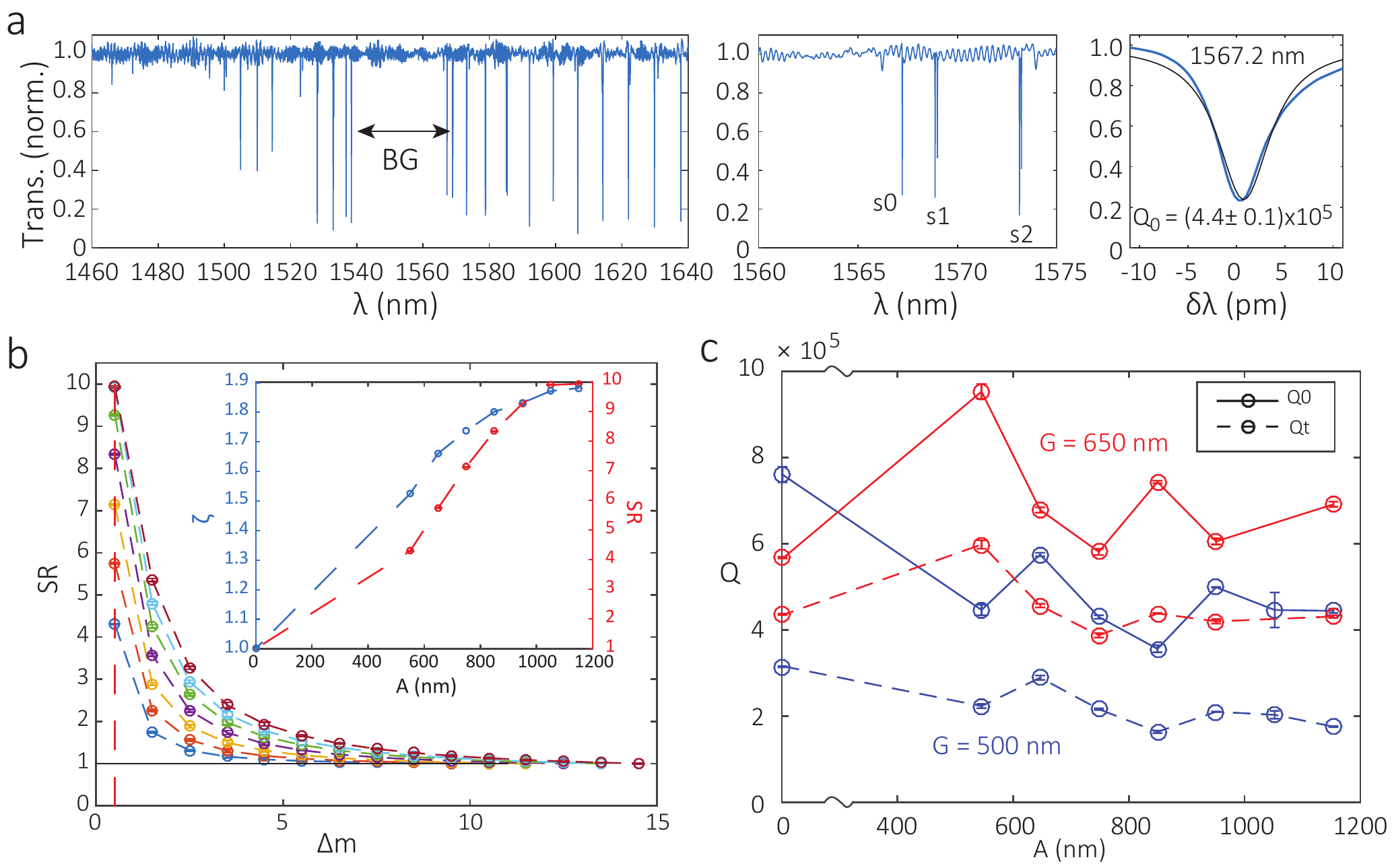}
\caption{\textbf{Supporting data for slow light \xlt{MPhCR}.} \textbf{a,} Characterization of a slow light device with $A$~=~550~nm modulation. This modulation opens up a 29~nm separation between modes in the middle of the spectrum. Comparing this device to the bare microring (Fig.~\ref{Fig2}(c)), we see that the air band and dielectric band are symmetrically located, which is atypical to most photonic crystal designs. The dielectric band, for example, shows a singlet resonance at the band-edge ($s0$), and doublet resonances for all other slow light modes ($s1$ and $s2$ shown here). The $s0$ mode in this device has a $SR$ of 4.29~$\pm$~0.01 (uncertainty propagated from the $FSR$ between $s0$ and $s1^{\pm}$, see Methods for details) and $Q_\text{0}$~=~(4.4$\pm$0.1)$\times$10$^5$ (uncertainty from one-standard deviation in nonlinear fitting). \textbf{b,} Extraction of $SR$ for devices with $A$ from 550~nm (blue) to 1150~nm (purple) using free spectral ranges (see Methods). Dashed lines are for guidance in viewing. The red vertical dashed line at $\Delta m$~=~0.5 is used to estimate the $SR$ of $s0$ mode, whose values are shown versus $A$ in the inset (red). The $SR$ is closely related to the `hyperbolic' index $\zeta$, the parameter describing the curvature of the dielectric band (blue). See Methods for the equation for $\zeta$. \textbf{c,} $Q$ analysis of the $s0$ modes of slow light devices for control devices ($A$~=~0~nm) and \xlt{MPhCR} devices with $A$ from 550~nm to 1150~nm. Devices with two waveguide-ring coupling gaps ($G$s) are shown with $G$~=~500~nm (blue) and $G$~=~650~nm (red). In general, there seems to exist parasitic loss for \xlt{MPhCR} devices, that is, extra loss (reduced $Q_\text{0}$ effectively) induced by coupling, in $G$~=~500~nm cases.}
\label{ExtFig3}
\end{figure*}

\begin{figure*}[t!]
\centering\includegraphics[width=1.00\linewidth]{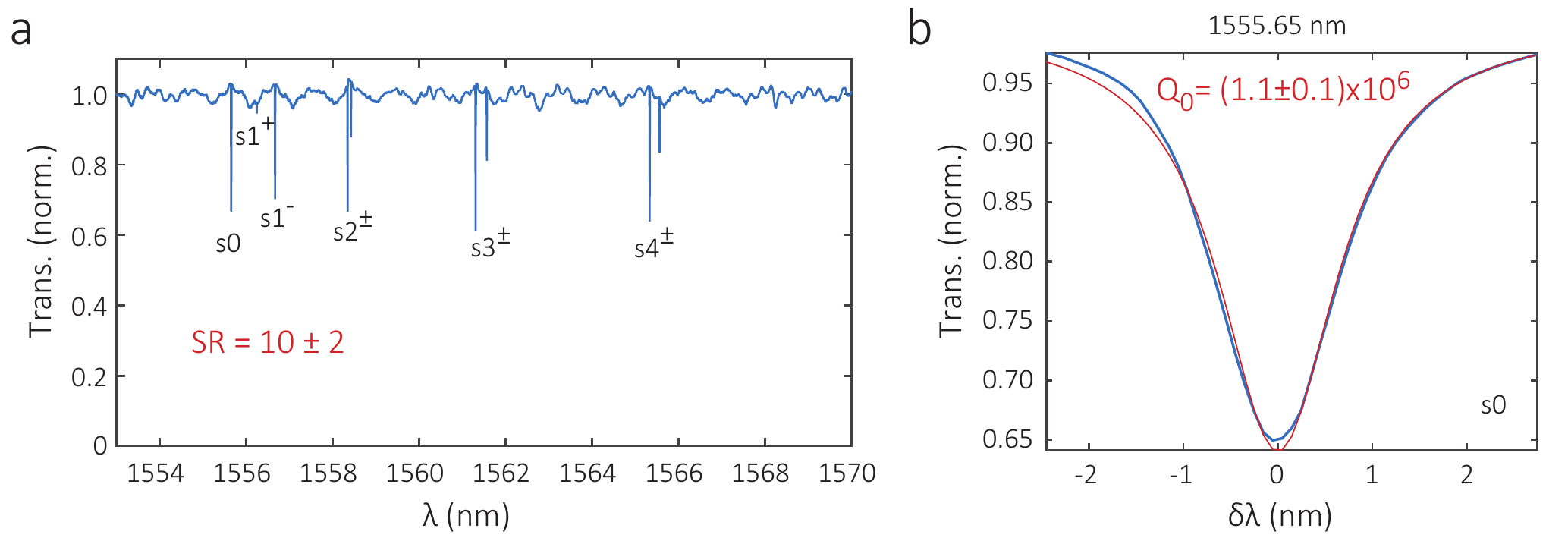}
\caption{\textbf{Supporting data on improving $Q$ of \kst{the MPhCR}.} \textbf{a,} A slow-light device after optimization shows $SR = 10 \pm 2$ for s0 and s1 modes. \textbf{b,} The s0 mode has an intrinsic optical $Q$ of 1.1 million.}
\label{ExtFig4}
\end{figure*}

\begin{figure*}[t!]
\centering\includegraphics[width=1.00\linewidth]{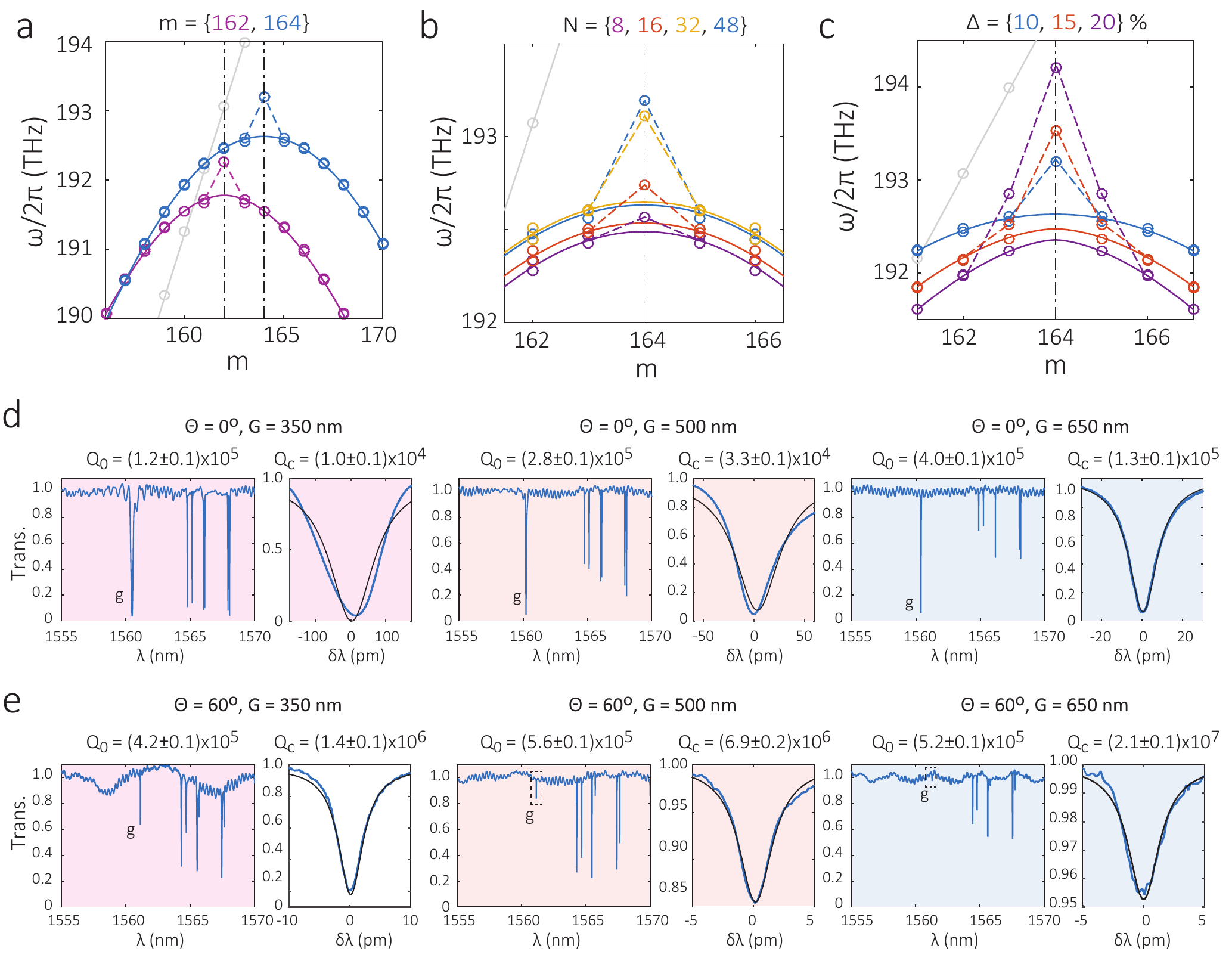}
\caption{\textbf{Supporting data on frequency control and coupling of dPhC modes in \kst{the MPhCR}.} \textbf{a-c,} Control of the frequency of the $g$ defect mode by a number of means besides varying the base modulation amplitude $A$ (as in Fig.~\ref{Fig3}(c)), including varying $m$ (the targeted azimuthal mode number), $N$ (the number of periods comprising the potential), and $\Delta$ (the modulation depth of the potential), from left to right, respectively. Nominal parameters are $\{$$m$, $N$, $\Delta$$\}$~=~$\{$162, 48, 10~\%$\}$. \kst{The MPhCR} has a unique tuning mechanism, in which the modulation period is varied to target different azimuthal mode numbers, for example, $m$~=~162 (blue) and $m$~=~164 (purple) in (a), while the characteristics of the defect ($\Delta$,$G$,and $N$) are unchanged. This changes the location of the band-edge, but not the curvature of the dielectric band and the relative frequency of the $g$ mode. This characteristic of the $g$ mode is particularly convenient considering the relation of \kst{the MPhCR} to control microrings (grey). Otherwise, tuning of the $g$ mode frequency by control of $N$ and $\Delta$ results in a largely unchanged dielectric band curvature, with its frequency shifted by $<$1~THz, while the $g$ mode frequency is tuned by as much as 2~THz. \textbf{d-e,} Coupling of the \xlt{MPhCR} devices and fitting of the $g$ modes for centered and rotated PhC defects. This series of transmission spectra shows the control of $Q_\text{c}$ from a regime of deep coupling ($Q_\text{c}\approx10^4$, $Q_\text{0}\approx10^5$, $K\approx10$) to one that is strongly undercoupled ($Q_\text{c}\approx2\times10^7$, $Q_\text{0}\approx5\times10^5$, $K\approx0.025$). We note that the $g$ modes have on-resonance transmission values close to 0 when $K~\gg~1$, unlike conventional coupling of microring traveling wave modes (see Methods).}
\label{ExtFig5}
\end{figure*}

\end{document}